\documentclass[prd,superscriptaddress,floatfix,amssymb]{revtex4}
\usepackage{epsf}
\newcommand{\picdir}[1]{./#1}     

\def\lesssim{\mathrel{\hbox{\rlap{\hbox{\lower4pt\hbox{$\sim$}}}\hbox{$<$}}}}
\def\gtrsim{\mathrel{\hbox{\rlap{\hbox{\lower4pt\hbox{$\sim$}}}\hbox{$>$}}}}
\begin{document}
\title{Nonlinear Inflaton Fragmentation after Preheating}
\date{\today}
\author{Gary N. Felder}
\affiliation{Department of Physics, Clark Science Center, 
Smith College Northampton, MA 01063, USA}
\author{Lev Kofman}
\affiliation{CITA, University of
Toronto, 60 St. George Street, Toronto, ON M5S 3H8, Canada}
\preprint{hep-ph/xxxxxxx}
\begin{abstract}
We consider the nonlinear dynamics of inflaton fragmentation during
and after preheating in the simplest model of chaotic inflation. While
the earlier regime of parametric resonant particle production and the
later turbulent regime of interacting fields evolving towards
equilibrium are well identified and understood, the short intermediate
stage of violent nonlinear dynamics remains less explored.  Lattice
simulations of fully nonlinear preheating dynamics show specific
features of this intermediate stage: occupation numbers of the scalar
particles are peaked, scalar fields become significantly non-gaussian
and the field dynamics become chaotic and irreversible. Visualization
of the field dynamics in configuration space reveals that nonlinear
interactions generate non-gaussian inflaton inhomogeneities with very
fast growing amplitudes. The peaks of the inflaton inhomogeneities
coincide with the peaks of the scalar field(s) produced by parametric
resonance.  When the inflaton peaks reach their maxima, they stop
growing and begin to expand. The subsequent dynamics is determined by
expansion and superposition of the scalar waves originating from the
peaks.  Multiple wave superposition results in phase mixing and
turbulent wave dynamics.  Thus, the short intermediate stage is
defined by the formation, expansion and collision of bubble-like field
inhomogeneities associated with the peaks of the original gaussian
field. This process is qualitatively similar to the bubble-like
inflaton fragmentation that occurs during tachyonic preheating after
hybrid or new inflation.
\end{abstract}
\maketitle

\section{Introduction}

The origin of matter in the universe from a decaying inflaton field in
the process of (p)reheating is a basic feature of all realistic
inflationary models. If four dimensional effective field theory is
sufficient (for specifics of reheating in string theory inflation see
e.g. \cite{KY}) this process is described by the non-equilibrium QFT
of particle creation and thermalization. In chaotic inflation this
particle creation typically involves a period of parametric resonance,
when occupation numbers of Bose particles rapidly become exponentially
large \cite{KLS}. In this case the QFT is well approximated by
classical field theory, which can be investigated in detail with
classical lattice simulations \cite{lattice}.  A full treatment of the
quantum field theory of the non-linear stages of preheating in
controllable models in the limit of high occupation numbers is in
agreement with the classical approximation \cite{berges}.

The simplest possible inflationary potential
contains a massive inflation field $V=\frac{1}{2} m^2 \phi^2$ and the simplest
preheating model involves a  coupling of the inflaton to another
field $ \frac{1}{2}g^2 \phi^2 \chi^2$.
 The regime of parametric resonant particle production
is  understood analytically  \cite{KLS}.
Backreaction of inhomogeneous fluctuations quickly brings the system 
of interacting scalar fields to a strongly  non-linear regime
characterized by very high occupation numbers.
The turbulent regime of interacting classical scalar field waves 
was studied in detail in numerical simulations \cite{lattice,MT,FK},
most of which are based on the LATTICEASY code \cite{FT}, and
in the $\lambda \phi^4$ model even analytically with the kinetic theory of Kolmogorov-type
turbulence \cite{MT}.
The least understood stage of (p)reheating is the short, violent transition from
linear preheating to the turbulent stage,
which shows  anomalies in the momentum space picture,
and in the departure from gaussian statistics \cite{FK}.

Hybrid inflation is another very important class of inflationary
models.  At first glance preheating in hybrid inflation, which
contains a symmetry breaking mechanism in the Higgs field sector, has
a very different character than in chaotic inflation.  Preheating in
hybrid inflation occurs via tachyonic preheating \cite{GBFKLT}, in
which a tachyonic instability of the homogeneous modes drives the
production of field fluctuations.  In hybrid inflation, the decay of
the homogeneous fields leads to fast non-linear growth of scalar field
lumps associated with the peaks of the initial (quantum)
fluctuations. The lumps then build up, expand and superpose in a
random manner to form turbulent, interacting scalar waves
\cite{GBFKLT,lumps}.  A similar picture emeres in new inflation
preheating \cite{new}.  Like parametric resonance, tachyonic
preheating can be interpreted via the reciprocal picture of copious
particle production far away from thermal equilibrium, and consequent
cascades of energy through interacting, excited modes.

In this paper we investigate in detail the structure of preheating
after chaotic inflation in configuration space. We find that the
intermediate stage between linear preheating and turbulence proceeds
via non-linear growth, expansion, and superposition of large value
field bubbles, similar to what we earlier observed in hybrid
inflation. Thus, the bubble-like intermediate structure of the
non-linear fields is another consistent feature of
preheating. However, the
details of the non-linear dynamics are different:
 while the bubbles in preheating after hybrid
inflation initially occur as isolated patches with large spaces in
between them, the bubbles that appear in preheating in the model being
considered here appear more densely throughout the space and persist
for some time as a pattern of standing waves before they begin
spreading and colliding.

In section \ref{model} we describe the model we are considering and
review the basic nature of preheating in this model, focusing on the
(well studied) behavior of the fields in momentum space. In section
\ref{results} we discuss different diagnostics of the system of
interacting scalar fields  in configuration space
and describe the fully non-linear dynamics of inflaton
fragmentation. In
the concluding section we discuss some of the implications of these
results, in particular, for the generation of gravitational waves from preheating
and baryo/leptogenesis from preheating.

\section{Preheating and Thermalization in Momentum Space}\label{model}

We consider the potential
\begin{equation}
\label{potential}
V = {1 \over 2} m^2 \phi^2 + {1 \over 2} g^2 \phi^2 \chi^2
\end{equation}
where $\phi$ is the inflaton and $\chi$ is another scalar field that
is coupled to it. At the end of inflation $\phi$ is a homogeneous
field which oscillates as $\phi(t) \approx \frac{M_p}{\sqrt{3\pi} mt} \, \sin
mt$ and $\chi$ is a quantum field with eigenfunctions $\chi_k(t) e^{-i{\bf k  x}}$.
 The temporal part $\chi_k(t)$ obeys an oscillator
equation with a periodic frequency $\omega_k^2=(k^2/a^2)+g^2
\phi(t)^2$.  The amplitude $\chi_k(t)$ thus undergoes parametric
resonance, leading to large occupation numbers of created particles
$n_k$.

Due to the rapid growth of its occupation numbers the field $\chi(t,
\vec x)$ can be treated as a classical scalar field.  Its appearance
is described by the realization of the random gaussian field
\begin{equation}
\label{field}
\chi(t \ ,  \vec x)= 2 \, \int d^3k \,   |\chi_k(t)| \, \cos ({\bf k  x}+\theta_k)
\end{equation}
i.e. as a superposition of standing waves with random phases
$\theta_k$ and Rayleigh-distributed amplitudes $P( |\chi_k(t)|)d
|\chi_k(t)|=e^{-\frac{|\chi_k(t)|^2}{\Delta}} \, \frac{2
|\chi_k(t)|}{\Delta} \, d|\chi_k(t)|$, $\Delta=<|\chi_k(t)|^2>$.  One
can use many different quantities to characterize a random field, such
as its variances $<\sigma^2_n>= \int d^3k \, k^{2n} \, |\chi_k(t)|^2$,
the spatial density of its peaks of a given height, etc. The scale of
the peaks and their density depend on the characteristic scale $R$ of
the spectrum, which in our case is related to the leading resonant
momentum $k_* \simeq \sqrt{g m \phi_0} a^{1/4}$ \cite{KLS}.  At the
linear stage the phases $\theta_k$ are constant, so that the structure
of the random field $\chi$ stays almost the same.

Once one field is amplified in this way, other fields that are coupled
to it are themselves amplified \cite{FK}, so within a short time of
linear preheating (of order dozens of inflaton oscillations)
fluctuations of $\chi$ generate inhomogenoeus fluctuations of the
field $\phi$.  It is easy to see that fluctuations of $\phi$ will have
a non-linear, non-gaussian character.  From the equation of motion for
$\phi$
\begin{equation}
\label{f1}
\Box \phi +m^2 \phi^2 +g^2 \chi^2 \phi=0 \ ,
\end{equation}
we have in Fourier-space
\begin{equation}
\label{fourier}
\ddot \phi_k+3H\dot \phi_k+\left( (k^2/a^2)+m^2\right)\phi_k=g^2 \phi_0(t) \, 
\int d^3q \, \chi_q \chi^*_{k-q} \ ,
\end{equation}
where we neglect the term that is third order with respect to
fluctuations; $\phi_0(t)$ is the background oscillation.
  The solution of this equation with Green's functions
\cite{KLS} shows that $\phi$ fluctuations grow with twice the exponent
of $\chi$ fluctuations. It also shows that the fluctuations of $\phi$ are non-gaussian.
Sometimes this solution is interpreted as re-scattering of
the particle $\chi_q$ against the condensate particle $\phi_0$ at rest
producing $\chi_{k-q}$ and $\phi_k$, $\chi \phi_0 \to \chi \delta \phi$.
As we will see shortly, this interpretation has significant limitations.

\begin{figure}[htb]
\begin{minipage}[t]{7.5cm}
\centering \leavevmode \epsfxsize=7.5cm
\epsfbox{\picdir{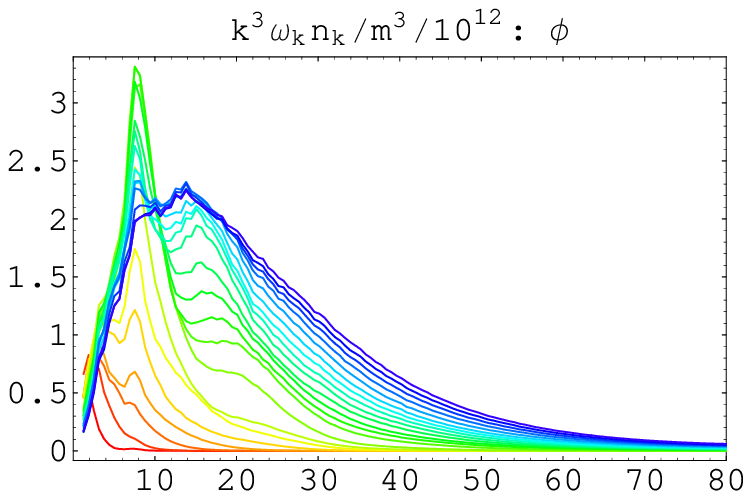}}\\
\end{minipage}
\begin{minipage}[t]{7.5cm}
\centering \leavevmode \epsfxsize=7.5cm
\epsfbox{\picdir{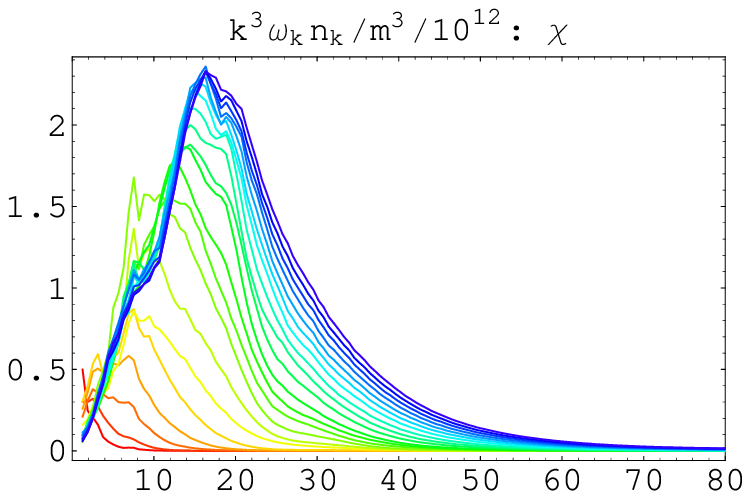}}\\
\end{minipage}
\caption{Evolution of spectra in the combination
$k^3 \omega_k n_k$
of the $\phi$ and $\chi$ fields during and
immediately after preheating. Bluer plots show later spectra.
Horizontal axis $k$ is in units of $m$}
\label{spectra}
\end{figure}

\begin{figure}[htb]
\begin{minipage}[t]{7.5cm}
\centering \leavevmode \epsfxsize=7.5cm
\epsfbox{\picdir{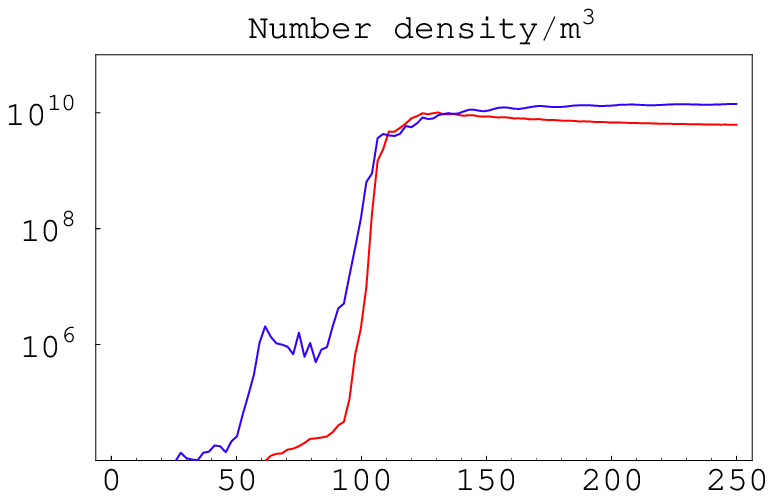}}\\
\caption{Evolution of comoving number density of $\phi$ (red, lower plot)
  and $\chi$ (blue, upper plot) in units of $m^3$. Time is in units of $1/m$}
\label{number}
\end{minipage}
\hspace{0.2cm}
\begin{minipage}[t]{7.5cm}
\centering \leavevmode \epsfxsize=7.5cm
\epsfbox{\picdir{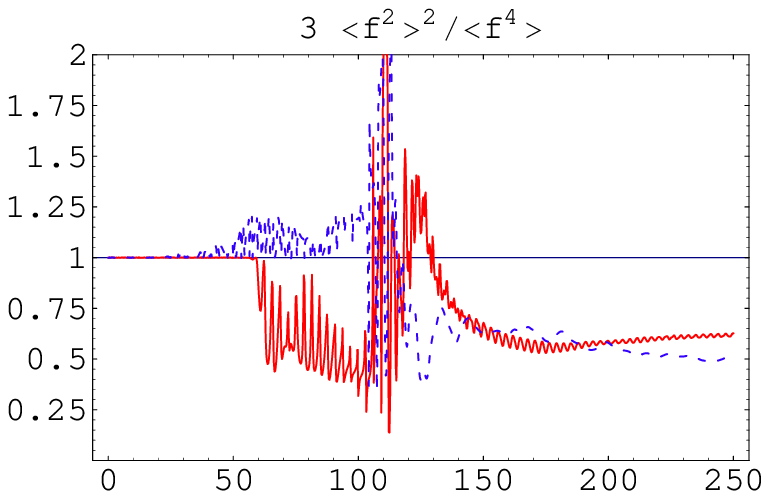}}\\
\caption{Evolution of the ratio $\langle f^2\rangle^2/\langle f^4\rangle$, where
$f$ represents the $\phi$ field (red, solid) or the $\chi$ field
(blue, dashed) and angle brackets represent a spatial average, is a
measure of gaussianity. This ratio is one for a random gaussian
field. Time is in units $1/m$}
\label{gaussian}
\end{minipage}
\end{figure}

When the amplitudes of $\chi$ and $\phi$ become sufficiently large we
have to deal with the fully non-linear problem.  The field evolution
can be well approximated using the classical equation of motion
(\ref{f1}) supplemented by another equation for $\chi$
\begin{equation}
\label{f2}
\Box \chi  +g^2 \phi^2 \chi =0 \ .
\end{equation}
Results of simulations of non-linear preheating using the LATTICEEASY
program \cite{FT} have been reported in many earlier papers \cite{lattice,FK}.
 For chaotic inflation, these results were presented in
terms of the time evolution of occupation numbers $n_k(t)$ or total
number density of particles $N(t)$. Figures
\ref{spectra}-\ref{gaussian} show the results of our simulations in
these familiar terms of $n_k(t)$ (in combination $k^3 \omega_k n_k$)
  and $N(t)$, as well as showing the
evolution of the field statistics (departures from gaussianity). Here
and for the rest of this paper all simulation results are for model
(\ref{potential}) with $m=10^{-6} M_p$ (fixed by CMB normalization)
and $g^2= 2.5 \times 10^{-7}$. The size of the box was $L=10
m^{-1}$ and the grid contained $256^3$ points. We also tried other
values of $g^2$ and found qualitatively similar results.

Figure \ref{spectra} shows the evolution of the spectra. The spectra
show rapid growth of the occupation numbers of both fields, with a
resonant peak that develops first in the infrared ($k \simeq k_*$) and
then moves towards the ultraviolet as a result of rescattering. In
figure \ref{number} you can see clearly that the occupation number of
$\chi$ initially grows exponentially fast due to parametric resonance,
followed by even faster growth of the $\phi$ field due to the
interaction, in accordance with the solution of eq (\ref{fourier}).

The gaussianity of classical fileds can be measured in different ways.
In Figure \ref{gaussian} we show the evolution of the ratio $\langle
f^2\rangle^2/\langle f^4\rangle$ (kurtosis), which is equal to unity
for a gaussian field.  During the linear stage of preheating, the
field fluctuations are a random gaussian field, see eq.~(\ref{field}),
reflecting the initial quantum fluctuations that seeded them.  The
inhomogenoeus field $\phi$ is generated as a non-gaussian field, in
agreement with the solution of eq (\ref{fourier}).  When the
fluctuation amplitude begins to get large, both fields are
non-gaussian. During the later turbulent stage both fields begin to
return to gaussianity.

Another known feature of preheating is the onset of chaos, when small
 differences in the initial conditions for the fields lead to
 exponentially divergent solutions: $D(t)\simeq e^{\lambda t}$, where
 $D$ is the distance in phase space between the solutions and
 $\lambda$ is the Lyapunov exponent (see \cite{FK} for details).  The
 distance $D$ begins to diverge exponentially exactly after the
 violent transition to the turbulent stage.

Let us summarize the picture which emerges when we study preheating,
turbulence and thermalization in momentum space with the occupation
numbers $n_k$.  There is initial exponential amplification of the
field $\chi$, peaked around the mode $k_*$. At this stage the $\chi$
fluctuations form a squeezed state, which is a superposition of
standing waves that make up a realization of a random gaussian
field. Interactions of the two fields lead to very rapid excitation of
fluctuations of $\phi$, with its energy spectrum also sharply peaked
around $k_*$. To describe generation of $\phi$ inhomogeneities, people
use the terminology of ``re-scattering''of waves.  However, there is a
short violent stage when occupation numbers have a sharply peaked and
rapidly changing spectrum. The field at this stage is non-gaussian,
which signals that the waves phases are correlated.  In some sense,
the concept of ``particles'' is not very useful around that time.  In
the later turbulent stage when $n_k(t)$ gradually evolves and
gaussianity is restored (due to the loss of phase coherency) the
picture of rescattering particles becomes proper.  As we will see in
the next section, gaussianity is not restored for some time after the
end of preheating.  To understand this violent, intermediate stage,
however, it is useful to turn to the reciprocal picture of field
dynamics in configuration space.

\section{Inflaton Fragmentation in Configuration Space}\label{results}

The features in the occupation number spectra $n_k(t)$, namely, sharp
time variations, peaks at $k \sim k_*$, and strong non-gaussianity of
the fields around the time of transition between preheating and
turbulence suggest that we dealing with distinct spatial features of
the fields in the configuration space.  This prompted us to study the
dynamics of the fields in configuration space.

\begin{figure}[htb]
\leavevmode\epsfxsize=.48\columnwidth\epsfbox{\picdir{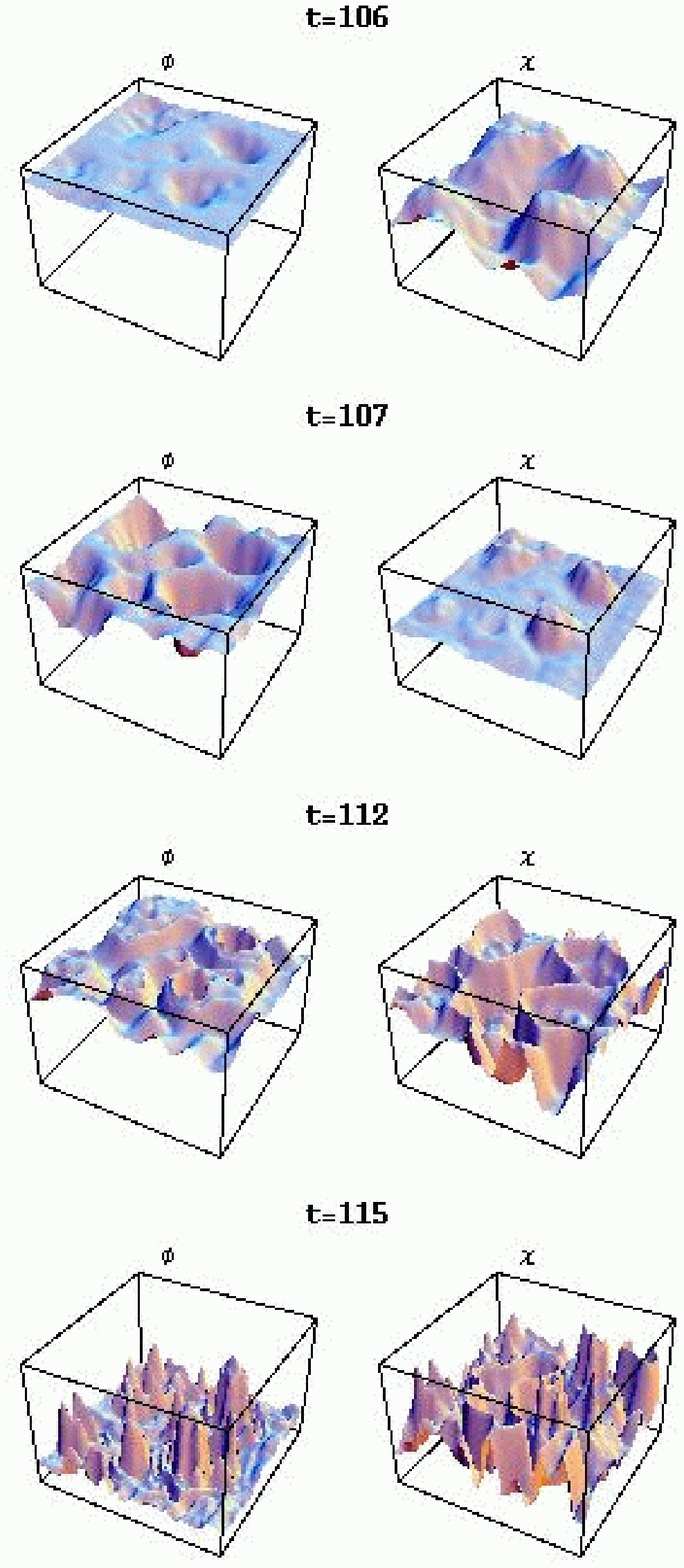}}
\hspace{.1cm}
\vline
\hspace{.1cm}
\leavevmode\epsfxsize=.48\columnwidth \epsfbox{\picdir{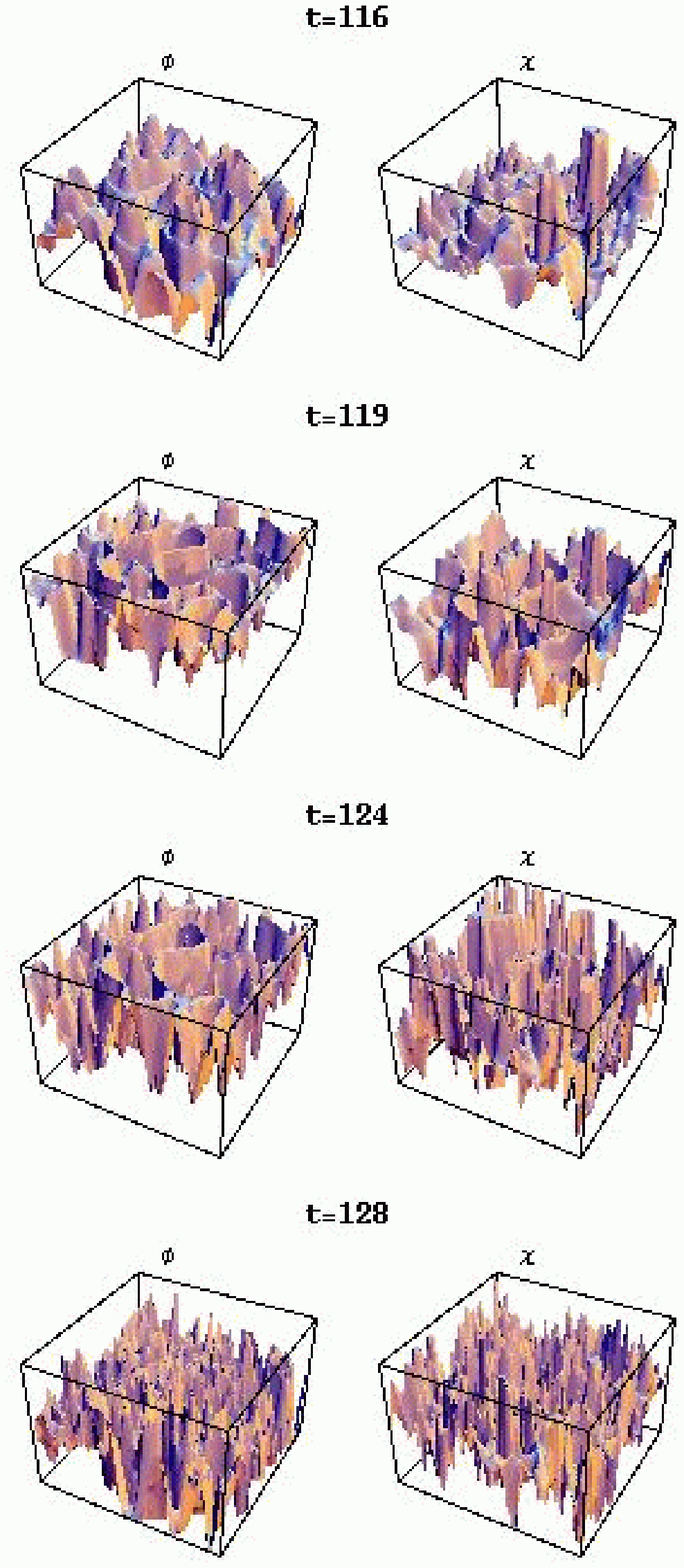}}
\caption{Values of the $\phi$ and $\chi$ fields in a two dimensional
slice through the lattice. The horizontal axes are spatial axes and
the vertical axis is field value.}
\label{3dslices}
\end{figure}

The evolution of the fields in configuration space is shown in Figure
\ref{3dslices}. Each frame shows the spatial profile of the fields
$\phi$ and $\chi$ along a two-dimensional slice of the 3D lattice. A
movie that includes many more time frames can be found at
{\tt
http://www.science.smith.edu/departments/Physics/fstaff/gfelder/public/bubbles/}.
Note that here (and everywhere in this paper) times are reported in
units of $1/m$.

The initial evolution of the fields ($t \lesssim 100$) is
characterized by linear growth of fluctuations of $\chi$. During this
stage the fluctuations have the form of a superposition of standing
waves with random phases, which make up a random gaussian field
(\ref{field}).  The eye captures positive and negative peaks that
correspond to the peaks of the initial gaussian random field $\chi$.
The peaks in this early stage correspond to the peaks of the initial
gaussian random field $\chi$. Next the oscillations of $\chi$ excite
oscillations of $\phi$. The first panel of Figure (\ref{3dslices})
shows a typical profile near the end of this period, just as the
oscillations are becoming nonlinear and $\phi$ is becoming
excited. The amplitude of these $\phi$ oscillations grows much faster
than the initial $\chi$ oscillations (see our discussion of
eq. (\ref{fourier})) and the oscillations have different (and changing)
frequencies.  The peaks of the $\phi$ oscillations occur in the same
places as the peaks of the $\chi$ oscillations, however, as can be
seen in the bottom three panels on the left side of Figure
(\ref{3dslices}). We can analyze the evolution during this early
non-linear period, not using the Fourier-mode description
(\ref{fourier}), but instead considering the configuration space
description equation (\ref{f1}). The interaction term in the equation
of motion is approximated during this stage by $g^2 \phi_0 \chi^2$, so
we have
\begin{equation}
\label{nonlin}
\phi(t, {\bf x})=\phi_0(t)+g^2 \int d^4x' \, \phi_0(t') \, \chi^2(t', {\bf x'}) G(s) \ , 
\end{equation}
where $G(s)$ is the retarded Green's function of the massive scalar
field wave equation, $s=(t-t')^2-({\bf x-\bf x'})^2$.  Thus the
profile of $\phi(t, {\bf x})$ is a superposition of the still
oscillating homogeneous part plus inhomogeneities induced by the
Yukawa-type interaction $(g^2 \phi_0) \phi \chi^2$ in the Lagrangian.
Since the Yukawa interaction is a short-range interaction (defined by
the length scale $1/m$ for a space-like interval $G(r) \sim e^{-mr}$),
induced inhomogeneities of $\phi$ appear in the vicinity of those in
$\chi$.

In the next stage ($t \gtrsim 110$) the peaks reach their maximum
amplitude, comparable to the initial value of the homogeneous field
$\phi$, and begin to spread. The two lower left panels of Figure
(\ref{3dslices}) shows the peaks expanding and colliding. In the
panels on the right you can see the standing wave pattern lose
coherence as the peaks send out ripples that collide and interfere. By
$t=124$ the fluctuations have spread throughout the lattice,
but you can still see waves spreading from the original locations of
the peaks. Shortly after that time all coherence is lost and the
field configurations appear to be like random turbulence.

\begin{figure}[htb]
\leavevmode\epsfxsize=.9\columnwidth\epsfbox{\picdir{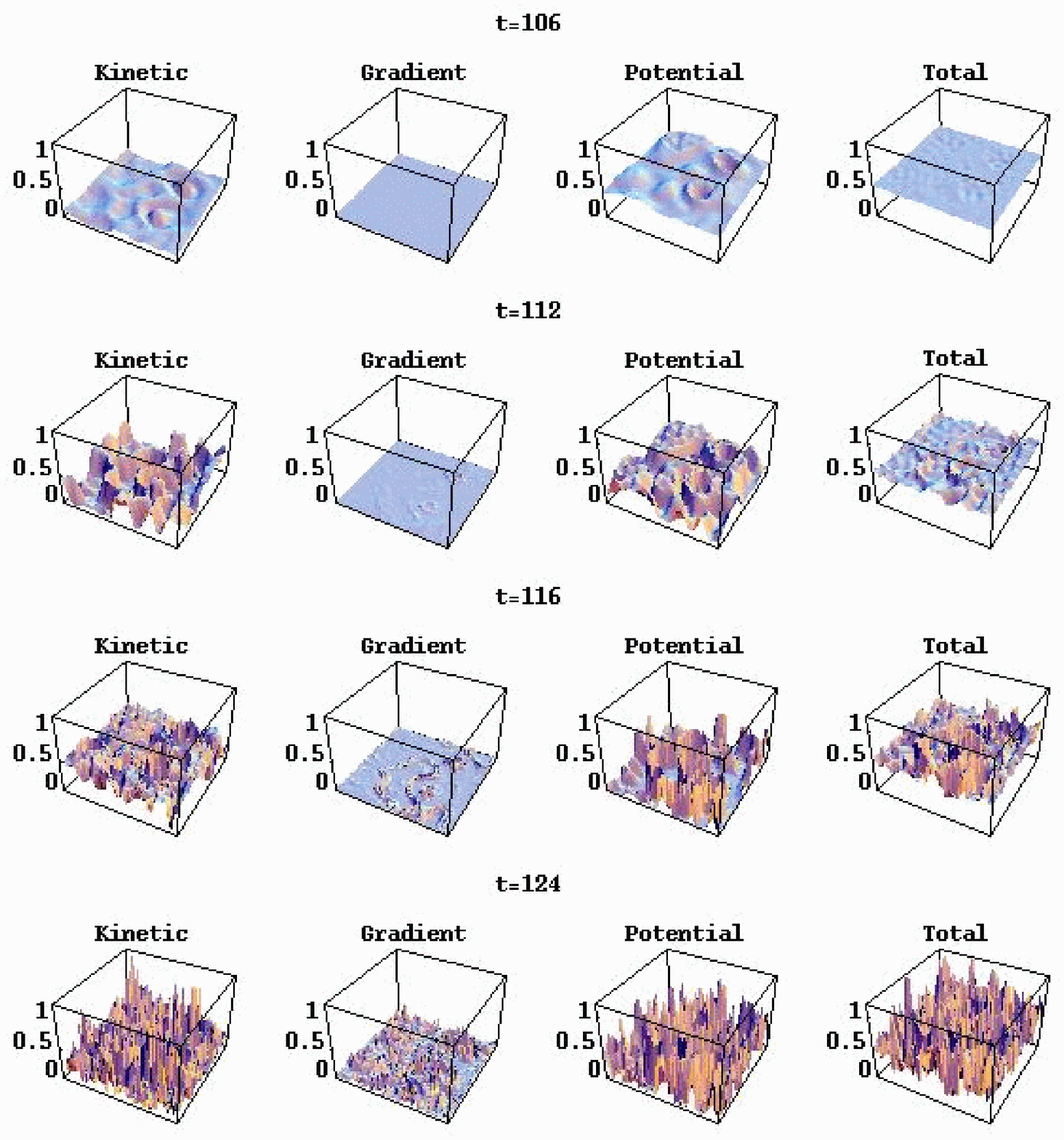}}
\caption{Energy density and its components. The horizontal axes
represent the same two dimensional slice through the lattice as in
figure (\ref{3dslices}).}
\label{energyslices}
\end{figure}

Figure (\ref{energyslices}) shows the distribution of energy density
at several points during the evolution. Several points about these
figures are worth noting. For the parameters we are considering, the
gradient energy is subdominant throughout the violent rescattering
stage and only begins to be significant during the onset of
turbulence. Fluctuations in the potential and kinetic energy grow in
the locations of the bubbles seen in figure (\ref{3dslices}), but they
are out of phase such that the total energy density remains nearly
homogeneous. Later, as the bubbles spread and collide, the phase
coherence is lost and inhomogeneities appear in the total energy
density. A movie showing many more time frames of energy density can
be found at {\tt
http://www.science.smith.edu/departments/Physics/fstaff/gfelder/public/bubbles/}.

\begin{figure}[htb]
\leavevmode\epsfxsize=.9\columnwidth\epsfbox{\picdir{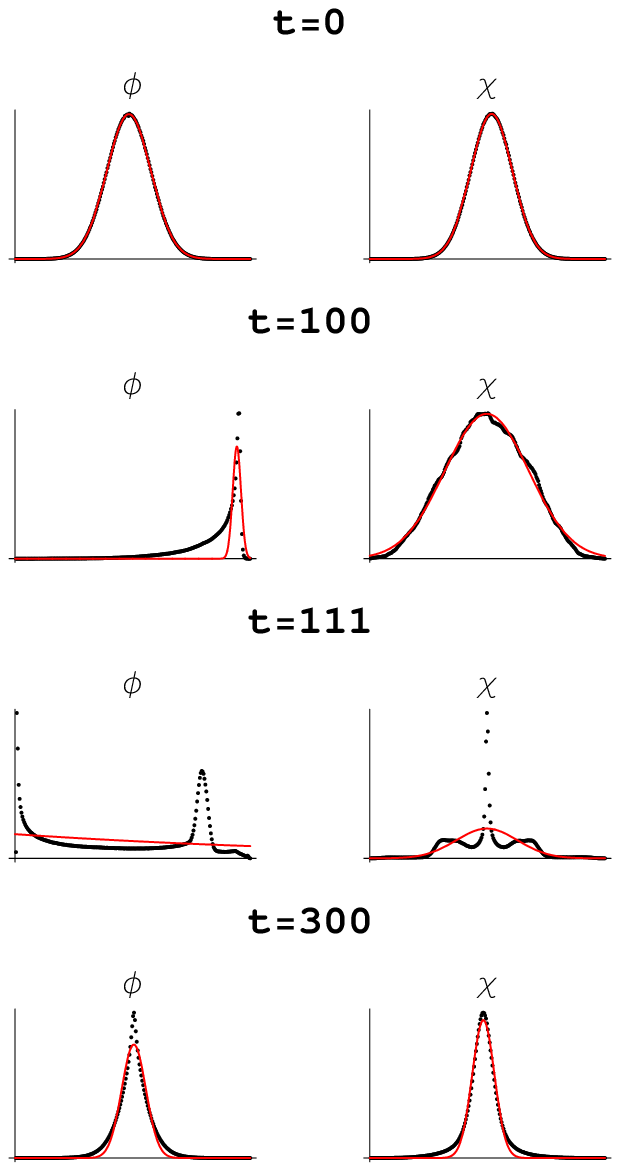}}
\caption{Field distributions. The horizontal axes are field values and
the vertical axis represents the frequency of that field value on the
lattice. The black dots represent the simulation results and the red
lines are best-fit Gaussian curves. The fitting lines can not be seen
in the first frame because they lie directly under the dots.}
\label{hist}
\end{figure}

For the first time we calculate the evolution of the statistics of the
fields. Figure (\ref{hist}) shows the field distributions (histograms)
at various times during the evolution.  Initially both fields have
gaussian distributions (from their random quantum fluctuations), with
$\phi$ sharply peaked around $\phi_0$ and $\chi$ centered around
zero. As non-linear effects become important, the statistics of both
fields become quite non-trivial. The distribution of the inflaton
field becomes at times sharply peaked to one side, when the condensate
is at an extreme end of its oscillation, and at other times bi-modal,
with a distinct presence of the homogeneous component plus a
significant inhomogeneous component.  The statistics of the $\chi$
field also become strongly non-gaussian.  Perhaps most surprisingly,
the statistics of both fields remain non-gaussian for a long time
after preheating. At the end of our simulation, at $t=300$, the fields
were still noticeably non-gaussian.  During all this time the random
phase approximation of interacting scalars is justified only to the
extent that the distributions approximate gaussians.

\section{Discussion}

It is convenient to split the process of transition from the
homogeneous inflaton condensate to the radiation of randomly moving
waves into four stages: The first is exponentially rapid growth of
small inhomogeneities that emerge from vacuum fluctuations. During
this stage fluctuations are linear and the fields are gaussian random
fields. The second stage is violent backreaction and rescattering of
waves, with non-gaussian, nonlinear, nonthermal fluctuations. The
third stage is Kolmogorov turbulence. During this period the
fluctuations are again (nearly) gaussian and energy gradually cascades
towards high momentum modes. Finally, there is thermalization.

In this paper we used lattice simulations to investigate the second
stage of violent field restructuring. We considered the model
(\ref{potential}), in which preheating occurs through parametric
resonance, and examined the evolution of the fields in configuration
space. The picture that emerged is similar in many ways to what we
observed earlier for tachyonic preheating in hybrid inflation.  For
instance in the F-term inflation example where the non-linear
potential around the bifurcation point is $V(\sigma)=-\frac{\lambda}{3}v
\phi^3 +\frac{\lambda}{4} \phi^4+\frac{\lambda}{12}v^4$,  small
initial random fluctuations of $\phi$ are amplified by the non-linear
$\phi^3$ term. The peaks of the gaussian field thus begin to grow very
fast relatively to the surrounding regions of $\phi$. Along those same
lines, we found here that oscillations of the $\chi$ and $\phi$ fields
grow initially at the locations of the peaks in the initial $\chi$
field. For the model (\ref{potential}) these growing peaks form a
pattern of standing waves that persists throughout the linear regime
and then begin to spread and overlap as rescattering becomes important.

The bubble-like structure we see here (and in hybrid inflation) has
nothing to do with first-order phase transitions, but just with
the initial structure of vacuum fluctuations plus non-linear dynamics
\footnote{Similar physics operates in a different context with
gravitational instability in an expanding universe, when growth of
initial matter inhomogeneities results in the cosmic web pattern of
large scale structure.  Clusters (voids) form from the high positive
(negative) peaks of the initial gaussian field, originated from vacuum
fluctuations during inflation. See e.g. J.R.~Bond, L.~Kofman,
D.~Pogosian, ``How filaments are woven into the cosmic web''.  Nature
380:603-606,1996.}.  Growth of the individual peaks results in the
build-up of the scalar field gradients.  Subsequent evolution is
defined by the expansion of the bubbles.  The superposition of many
almost spherically expanding bubbles leads to decoherence and
turbulent motion of the scalar waves.

The main lesson we have learned from this work is that the preheating
stage of linear fluctuations and the turbulent stage of interacting
waves are divided by a short, violent stage of non-linear formation
and collision of bubble-like large value field regions.

Eventually the fields reach thermal equilibrium characterized only by
the temperature. Does that mean that all traces of inflaton
fragmentation history are erased?  There are potential tracers of
the non-linear stage of preheating related to out-of-equilibrium
processes. For instance, people have discussed realizations of
baryogenesis at the electroweak scale via tachyonic preheating after
hybrid inflation \cite{GGKS}, and this process is ultimately related
to the bubble-like lumps of the Higgs field that form during tachyonic
preheating \cite{GGG}. Since we now see that fragmentation through
bubbles can also occur in chaotic inflation, baryogenesis via
out-of-equilibrium bubbles can also be extended to these models.

There is another, potentially observable consequence of the non-linear
``bubble'' stage of inflaton fragmentation. Lumps of the scalar fields
correspond to large (order of unity) energy density inhomogeneities at
the scale of those bubbles, $R$. Collisions of bubbles generate
gravitational waves.  The fraction of the total energy at the time of
preheating converted into gravitational waves is significant. We
estimate it is of the order of
\begin{equation}\label{fraction}
\frac{\rho_{gw}}{\rho_{rad}} \simeq (RH)^2 \ ,
\end{equation}
where $1/H$ is the Hubble radius. This corresponds to a present-day
fraction of energy density $\Omega_{GW} \sim 10^{-5}(RH)^2$. 
The way to understand formula (\ref{fraction})  is the following:
The energy converted into gravitational waves from the collision of
two black holes is of the order of the black hole masses. If the mass
of lumps of size $R$ is a fraction $f$ of a black hole of the same
size, then the fraction of energy converted to gravitational waves
from two lumps colliding is $f$. Scalar field lumps at the Hubble
scale would form black holes, so in our case $f=(RH)^2$.

The
present-day frequency of this gravitational radiation is
\begin{equation}\label{frequency} 
f\simeq \frac{10^7 Gev}{M} Hz \ ,
\end{equation}
where $M=V^{1/4}$ is the energy scale of inflation with the potential
$V$.

For the chaotic inflation model considered in this paper the size of
the bubbles is $R \sim few/m$ and at the time they begin colliding $H
\sim m/100$, so that the fraction of energy converted into
gravitational waves is of the order $10^{-3}-10^{-4}$.  This figure is
in agreement with the numerical calculations of gravitational wave
radiation from preheating after chaotic inflation \cite{GW}.

For chaotic inflation with $M$ at the GUT scale the frequency
(\ref{frequency}) is too short and not observable.  Gravitational
waves continue to be generated during the turbulent stage and even
during equilibrium due to thermal fluctuations, but with a smaller
amplitude. It is a subject of further investigation if they can be
observed.  The most promising possibility for observations is,
however, generation of gravity waves from low energy hybrid inflation,
where $f$ can much much smaller.

\bigskip

We would like to thank Marco Peloso for useful discussions. G.F. was
supported by NSF grant PHY-0456631.
L.K. wa supported by NSERC and CIAR. G.F. would like to thank the
Canadian Institute for Theoretical Astrophysics for their hospitality
during part of this work

\end{document}